\documentclass[a4paper, 10pt, conference]{ieeeconf}      

\IEEEoverridecommandlockouts                              

\usepackage{graphicx}
\usepackage{amsmath}
\usepackage{multirow}
\usepackage{graphicx}
\usepackage{threeparttable}
\usepackage{tikz}
\usepackage{dsfont}
\usepackage{url}

\newcommand{\R}{\mathds{R}}
\newcommand{\E}{\mathds{E}}

\begin{document}

\title{\LARGE Data Augmentation of IMU Signals and Evaluation via a Semi-Supervised Classification of Driving Behavior}

\author{Amani Jaafer$^1$\thanks{$^1$A.~Jaafer is with the Division of Systems Analysis and Economics, Royal Institute of Technology (KTH), Stockholm, Sweden. \texttt{jaafer@kth.se}}, 
Gustav Nilsson$^2$\thanks{$^2$G.~Nilsson is with the School of Electrical and Computer Engineering, Georgia Institute of Technology, Atlanta, GA, USA. \texttt{gustav.nilsson@gatech.edu}}, and  Giacomo Como$^3$\thanks{$^3$G.~Como is with the Department of Mathematical Sciences, Politecnico di Torino, Italy and the Department of Automatic Control, Lund University, Sweden. \texttt{giacomo.como@polito.it}}\thanks{This work was partially supported through research cyberinfrastructure resources and services provided by the Partnership for an Advanced Computing Environment (PACE) at the Georgia Institute of Technology, Atlanta, Georgia, USA, as well as by MIUR grant Dipartimenti di Eccellenza 2018--2022 [CUP: E11G18000350001], the Swedish Research Council [2015-04066], and Compagnia di San Paolo.}}

\maketitle
\thispagestyle{empty}
\pagestyle{empty}

\begin{abstract}

Over the past years, interest in classifying drivers' behavior from data has surged. Such interest is particularly relevant for car insurance companies who, due to privacy constraints,  often only have access to data from Inertial Measurement Units (IMU) or similar. In this paper, we present a semi-supervised learning solution to classify portions of trips according to whether drivers are driving aggressively or normally based on such IMU data. Since the amount of labeled IMU data is limited and costly to generate, we utilize Recurrent Conditional Generative Adversarial Networks (RCGAN) to generate more labeled data. Our results show that, by utilizing RCGAN-generated labeled data, the classification of the drivers is improved in $79\%$ of the cases, compared to when the drivers are classified with no generated data.
\end{abstract}

\section{Introduction}
Modern vehicles are equipped with an increasing number of sensing devices, such as Global Positioning System (GPS), Inertial Measurement Units (IMU), and other sensors that communicate through the Controller Area Network (CAN-Bus). This real-time sensed data can be used to detect, analyze, predict, and plan a large variety of issues such as traffic congestion, vehicle energy consumption and emissions, urban mobility, and drivers' behavior. Multiple approaches have been developed and applied to accurately identify driving behavioral patterns, such as driver recognition~\cite{choi2007analysis,mcnew2012predicting}, maneuver recognition~\cite{oliver2000driver,brambilla2017comparison,enev2016automobile}, and aggressive driving detection~\cite{carmona2015data}. While an accurate classification of the driving behavior  can contribute to a better driving experience for the driver, there are also other applications where such classification can be useful.

Recently, there are been a growing interest from car insurance companies in designing driver behavior classification systems that could eventually be used to relate their costumers' fees to how they drive. As a part of this solution, it is of interest to accurately classify the level of aggressiveness of their customers' recorded trips. Nevertheless, the large number of trips would not allow to identify for each one the type of driving. Consequently, several works such as~\cite{brambilla2017comparison}, \cite{fugiglando2018driving}, and \cite{8590215} have been conducted to solve this problem by an unsupervised learning approach. In the mentioned work, the goal is to find clusters from the recorded trip data which can be characterised by different levels of the aggressiveness without relying on the labels.

Since the labels, i.e., the driving style, remain a crucial element in order to apply a supervised algorithm, generating realistic artificial data can be an alternative to increase the size of the training or validation datasets and possibly improve the quality of the classification. Semi-supervised learning is motivated by the availability of large datasets with unlabeled features in addition to labeled ones, in different applications \cite{bahi2018deep}, \cite{Raina:2007:SLT:1273496.1273592}, \cite{zhang2009mining}. This lack of labeled data can be efficiently addressed through a deep learning pipeline.

Another application of interest for driving behavior classification is the development of autonomous vehicles. A better understanding of how humans drive can indeed allow for both a better functioning on a technical level and, of course, minimizing as much as possible any error, in view of the security of the users. Identifying aggressive drivers is crucial in developing safer autonomous driving techniques and advanced driving assistant systems. This problem has been extensively studied over the past decades in several works  \cite{wei2014behavioral}, \cite{horii2017modifying}, \cite{al2001framework}, \cite{bojarski2016end}.
Current autonomous driving systems use a wide range of algorithms to process sensor data. Some work, as~\cite{Cheung2018ClassifyingDB}, uses end-to-end approaches to make navigation decisions from the sensor inputs such as camera images, LIDAR data, etc. A variety of sensors can be useful for cars to extract important information to improve the quality of autonomous vehicles and to learn how to drive safely and efficiently. Nevertheless, data collection can also be expensive and restricted in terms of privacy. Simulating data and exploiting it in the same context as real ones appears as a solution to study. The attention to generative models is increasing due to their capability of modelling underlying patterns in multidimensional data. However, assessing the quality of the synthetic data remains a crucial point to validate. 

In this paper, we formulate the problem of generating labeled IMU signals, representing aggressive and normal drivers, of one-minute length for a specific part of the road, using Recurrent Conditional GANs. The generated data will be practically assessed based on its capacity to improve the classification of the semi-supervised framework. 

\section{Related work}
Since obtaining real sensor data can be costly, time-consuming, and have privacy issues, there have recently been several studies on sensor modelling for virtual testing, e.g. in~\cite{hirsenkorn2016virtual}, \cite{bernsteiner2015radar}, \cite{wheeler2017deep}, \cite{hirsenkorn2017learning} which are mostly based on parametric models. In~\cite{hirsenkorn2016virtual}, a non-parametric statistical model was developed allowing for the generation of sensor position output. In~\cite{bernsteiner2015radar} a radar model is proposed where noise is added to the raw signals, and then filtering is applied to model sensor output. Further, \cite{wheeler2017deep} proposed a Variational Autoencoder (VAE) approach in order to model the radar sensor output given some input vector, using object lists and spatial rasters. In \cite{zec2018statistical}, an Autoregressive InputOutput Hidden Markov Model (AIO-HMM) was proposed by fusing sensory streams through a linear transformation of features to synthesize real-valued time series describing sensor errors based on data describing the environment.

Generative Adversarial Networks (GANs)~\cite{goodfellow2014generative} have proven to perform well in generating different types of data. Different research works, from computer vision~\cite{bansal2018recycle}, \cite{dai2017towards}, \cite{gupta2018social}, to natural language processing \cite{xu2017neural}, had shown that the application of this kind of generative models can provide good results. In~\cite{Arnelid2018SensorMW} a Recurrent Conditional Generative Adversarial Network (RCGAN) has been proposed for modelling real-valued time series describing sensor outputs that are used in autonomous driving applications. In~\cite{sallab2019lidar}, the authors augmented the LiDAR sensor data in simulated environments, by employing CycleGANs.

Evaluating GANs is a challenging task. Unlike other deep learning models which are trained with a loss function until convergence, a GAN generator is trained and combined with a discriminator that learns to distinguish between the real or fake data. Both the generator and discriminator model are trained together to reach an equilibrium. Hence, there is no objective loss function used to train separately the GAN generator models. 
Some would rely on a  visual assessment by having appealing results that agree with the real distribution. The latter shows a high potential for some data, especially for images. Meanwhile, when time series are inspected visually, this remains an inconsistent method, since it is based on a manual operation to inspect each generated sample. By evaluating a convenient distance metric between the real and fake data distribution, we can assess the trained model and infer how much the model is capturing temporal patterns. Quantitative measures, such as reconstruction loss, Kullback-Leibler (KL) divergence, Jensen-Shannon (JS) divergence can be combined with visual assessment to provide a robust assessment of GAN models. A quantitative extrinsic approach like in~\cite{sallab2019lidar} and~\cite{esteban2017real} is also an alternative, which mainly relies on an external method to measure the quality of the generated data.

\section{Contribution}
This paper makes two main contributions to the field of driving behavior classification. First, it addresses the problem of data augmentation of car sensors. In our study, we generate IMU signals of one-minute length in a common portion of the road characterised by the type of driving style, which is either aggressive or normal. We use Recurrent Conditional GANs for the generation of these labeled time series. Second, we build a framework to evaluate the quality of generated data from a practical perspective. In other words, we assess the quality of the data based on the improvement of a semi-supervised model, which identifies the type of driving, by adding different percentage of synthesised data to the classifier's training and/or validation sets. Consequently, the paper investigates how much data should be generated and in which set should be used, to improve the accuracy of the driving behavior's classification.

\section{Approach}
\begin{figure} 
  \centering
  \includegraphics[width=0.85\linewidth, trim=0 150 0 0]{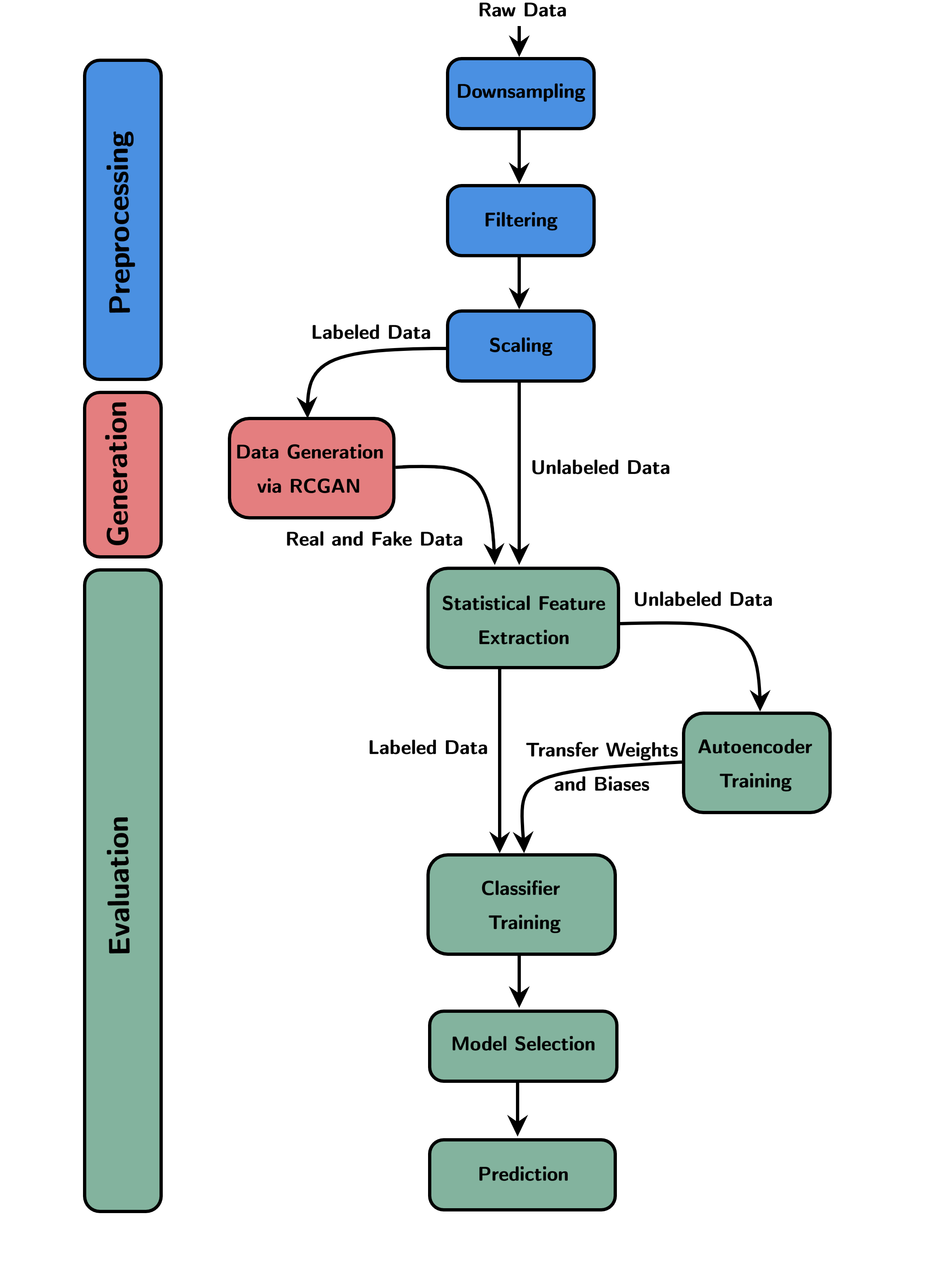}
  \caption{The overall framework used for classifying the drivers, including data preprocessing, data generation and evaluation through the semi-supervised approach.}
  \label{fr} 
\end{figure}

In this section, we firstly present the experimental setting used to collect the labeled data. Then, we present the data preprocessing tasks, followed by the generative model used to synthesize the multidimensional time series. Finally, we present the extrinsic assessment framework used to evaluate the generated data. The entire pipeline is shown in Fig.~\ref{fr}.

\subsection{Experimental Setting}
\begin{figure}
\centering
\includegraphics[width=6cm, trim=0 20 0 0]{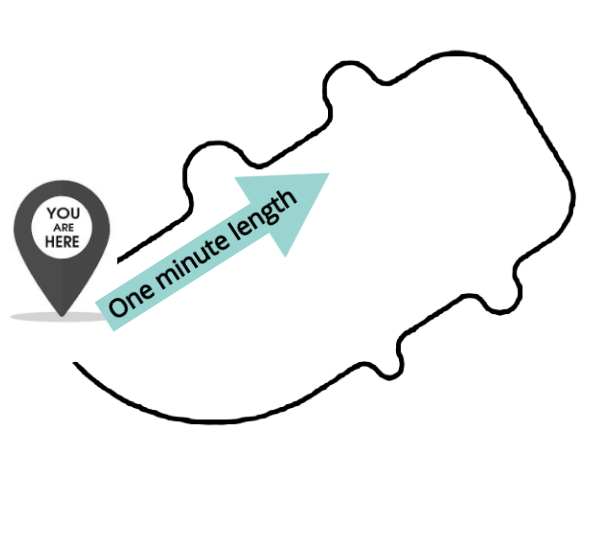}
\caption{The circuit the drivers where driving in the car simulator.}
\label{fig:circut}
\end{figure}

The dataset being used in this paper was collected from a vehicle simulator. The experiment consisted of $40$ drivers driving separately using different cars, in the same circuit. 
The circuit is depicted in Fig.~\ref{fig:circut}. 
The drivers had been asked to drive both in a normal and in an aggressive way. By doing so, we have close to a ground-truth about which recorded trips that are normal or aggressive. The simulator was collecting the same signals as a real IMU unit, i.e., longitudinal acceleration, lateral acceleration, pitch, yaw,  and roll. All the signals had the same sampling frequency, namely 1000~Hz. In total, the dataset consists of $n=238$ simulation drives.

\subsection{Data preprocessing}
For computational reasons, we down-sampled the IMU signals to 1 Hz, by taking each $1000^\text{th}$ observation. Although this down-sampling is done mostly for computational convenience, it is also very likely that in practical applications the hardware will have a more limited sampling frequency compared to the simulator. Since the signals may contain artifacts, we filtered them by applying a moving average filter with a sliding window of ten samples. We limited our study only on the first one minute of each trip, both for computational reasons, but also since in practical applications, it would be favorably to classify the driver without too much history. A similar choice time-window has previously been suggested in~\cite{carmona2015data}.

 All features were normalized using a MinMax scaler ~\cite{patro2015normalization}. Our dataset was split into the labeled data used in training the RCGAN and the unlabeled one used in the semi-supervised part. The RCGAN was trained only on a dataset of $60$ trips, equally balanced between aggressive and normal.

\subsection{Data generation}
RCGANs were originally developed and implemented in~\cite{esteban2017real} for medical applications. Our paper was inspired from this work to synthesize IMU signals for a normal and an aggressive trip. We will start this section by give a brief introduction to Recurrent Neural Networks (RNN). Next, we will introduce long short-term memory RNNs, which is an extension of the RNN framework. This subsection ends with a description of the RCGAN model, which is using long short-term memory RNNs.

\subsubsection{Recurrent Neural Networks}
RNNs are mostly used for sequential modeling and learning. They process one element of input
data at a time $t$ and implicitly store previous information using cyclic connections of hidden units. Given a sequence of vectors, $x = (x_{1},...,x_{T})$, where $x_{t} \in \R^{d_\text{in}}$, the RNN outputs a representation, that is a sequence of vectors $h = (h_{1},..., h_{T})$, where $h_{t} \in \R^{H}$. The sequence $h$ is determined iteratively through: 
\begin{equation}
    h_{t} = g(Wx_{t} + Uh_{t-1} + b) \,, \quad \forall t \in \{1,\ldots, T\}\,,
\end{equation}
where  $W \in \R^{H \times d_\text{in}}, U \in \R^{H \times H}, b \in \R^{H}$, and $h_0 = \mathbf{0}$. The function $g$ is a non-linear mapping and often chosen as $\tanh$ applied component-wise.

The output vector $p_{t} \in \R^{d_\text{out}}$ transforms the current hidden state $h_{t} \in \R^{H}$ in a way that
depends on the final task. For classification, it is computed as
\begin{equation}
    p_{t} = \text{softmax}(W_{p}h_{t} + b_{p}) \,.
\end{equation}
Note that $W \in \R^{H \times d_{in}}, U \in \R^{H \times H}, b \in \R^{H}, W_{p} \in \R^{d_\text{out} \times H}, b_{p} \in \R^{d_\text{out}}$ are network parameters determined through gradient descent. The scalars $H, d_\text{in}$ and $d_\text{out}$ are the dimensions of the hidden layer, the input, and the output, respectively. For example, in the case of 2-category classification, $d_\text{out} = 2$ and the probability vector $p_{t}$ refers to the probabilities of each input element $x_{t}$ belonging to each category.

\subsubsection{Long Short-Term Memory (LSTM)}
\begin{figure}
\centering
\tikzset{every picture/.style={line width=0.75pt}} 
\begin{tikzpicture}[x=0.75pt,y=0.75pt,yscale=-1,xscale=1,scale=0.8]
\draw  [line width=1.5]  (122.23,47.6) -- (395.84,47.6) -- (395.84,220.37) -- (122.23,220.37) -- cycle ;
\draw  [line width=1.5]  (247.87,128.43) .. controls (247.87,119.17) and (255.93,111.66) .. (265.88,111.66) .. controls (275.82,111.66) and (283.88,119.17) .. (283.88,128.43) .. controls (283.88,137.7) and (275.82,145.2) .. (265.88,145.2) .. controls (255.93,145.2) and (247.87,137.7) .. (247.87,128.43) -- cycle ;
\draw  [line width=1.5]  (122.59,128.43) .. controls (122.59,119.17) and (130.65,111.66) .. (140.59,111.66) .. controls (150.53,111.66) and (158.59,119.17) .. (158.59,128.43) .. controls (158.59,137.7) and (150.53,145.2) .. (140.59,145.2) .. controls (130.65,145.2) and (122.59,137.7) .. (122.59,128.43) -- cycle ;
\draw  [line width=1.5]  (304.4,128.77) .. controls (304.4,119.51) and (312.46,112) .. (322.4,112) .. controls (332.34,112) and (340.4,119.51) .. (340.4,128.77) .. controls (340.4,138.03) and (332.34,145.54) .. (322.4,145.54) .. controls (312.46,145.54) and (304.4,138.03) .. (304.4,128.77) -- cycle ;
\draw  [line width=1.5]  (247.87,203.6) .. controls (247.87,194.33) and (255.93,186.82) .. (265.88,186.82) .. controls (275.82,186.82) and (283.88,194.33) .. (283.88,203.6) .. controls (283.88,212.86) and (275.82,220.37) .. (265.88,220.37) .. controls (255.93,220.37) and (247.87,212.86) .. (247.87,203.6) -- cycle ;
\draw  [line width=1.5]  (168.67,64.7) .. controls (168.67,55.44) and (176.73,47.93) .. (186.67,47.93) .. controls (196.61,47.93) and (204.67,55.44) .. (204.67,64.7) .. controls (204.67,73.97) and (196.61,81.48) .. (186.67,81.48) .. controls (176.73,81.48) and (168.67,73.97) .. (168.67,64.7) -- cycle ;
\draw  [line width=1.5]  (179.83,129.1) .. controls (179.83,125.21) and (183.22,122.06) .. (187.39,122.06) .. controls (191.57,122.06) and (194.95,125.21) .. (194.95,129.1) .. controls (194.95,132.99) and (191.57,136.15) .. (187.39,136.15) .. controls (183.22,136.15) and (179.83,132.99) .. (179.83,129.1) -- cycle ;
\draw  [line width=1.5]  (258.31,157.95) .. controls (258.31,154.06) and (261.7,150.91) .. (265.88,150.91) .. controls (270.05,150.91) and (273.44,154.06) .. (273.44,157.95) .. controls (273.44,161.84) and (270.05,164.99) .. (265.88,164.99) .. controls (261.7,164.99) and (258.31,161.84) .. (258.31,157.95) -- cycle ;
\draw  [line width=1.5]  (369.92,129.1) .. controls (369.92,125.21) and (373.31,122.06) .. (377.48,122.06) .. controls (381.66,122.06) and (385.04,125.21) .. (385.04,129.1) .. controls (385.04,132.99) and (381.66,136.15) .. (377.48,136.15) .. controls (373.31,136.15) and (369.92,132.99) .. (369.92,129.1) -- cycle ;
\draw  [line width=1.5]  (358.76,64.7) .. controls (358.76,55.44) and (366.82,47.93) .. (376.76,47.93) .. controls (386.7,47.93) and (394.76,55.44) .. (394.76,64.7) .. controls (394.76,73.97) and (386.7,81.48) .. (376.76,81.48) .. controls (366.82,81.48) and (358.76,73.97) .. (358.76,64.7) -- cycle ;
\draw [line width=1.5]    (159.31,129.1) -- (175.83,129.1) ;
\draw [shift={(179.83,129.1)}, rotate = 180] [fill={rgb, 255:red, 0; green, 0; blue, 0 }  ][line width=0.08]  [draw opacity=0] (13.4,-6.43) -- (0,0) -- (13.4,6.44) -- (8.9,0) -- cycle    ;
\draw [line width=1.5]    (195.67,129.1) -- (242.79,128.79) ;
\draw [shift={(246.79,128.77)}, rotate = 539.62] [fill={rgb, 255:red, 0; green, 0; blue, 0 }  ][line width=0.08]  [draw opacity=0] (13.4,-6.43) -- (0,0) -- (13.4,6.44) -- (8.9,0) -- cycle    ;
\draw [line width=1.5]    (283.88,128.43) -- (300.4,128.7) ;
\draw [shift={(304.4,128.77)}, rotate = 180.94] [fill={rgb, 255:red, 0; green, 0; blue, 0 }  ][line width=0.08]  [draw opacity=0] (13.4,-6.43) -- (0,0) -- (13.4,6.44) -- (8.9,0) -- cycle    ;
\draw [line width=1.5]    (340.4,128.77) -- (365.92,129.06) ;
\draw [shift={(369.92,129.1)}, rotate = 180.65] [fill={rgb, 255:red, 0; green, 0; blue, 0 }  ][line width=0.08]  [draw opacity=0] (13.4,-6.43) -- (0,0) -- (13.4,6.44) -- (8.9,0) -- cycle    ;
\draw [line width=1.5]    (385.04,129.1) -- (412.11,129.33) ;
\draw [shift={(416.11,129.37)}, rotate = 180.49] [fill={rgb, 255:red, 0; green, 0; blue, 0 }  ][line width=0.08]  [draw opacity=0] (13.4,-6.43) -- (0,0) -- (13.4,6.44) -- (8.9,0) -- cycle    ;
\draw  [color={rgb, 255:red, 0; green, 0; blue, 0 }  ,draw opacity=1 ][fill={rgb, 255:red, 0; green, 0; blue, 0 }  ,fill opacity=1 ][line width=1.5]  (184.84,129.22) .. controls (184.86,128.02) and (185.92,127.05) .. (187.21,127.07) .. controls (188.5,127.08) and (189.54,128.07) .. (189.52,129.27) .. controls (189.51,130.48) and (188.45,131.44) .. (187.16,131.43) .. controls (185.87,131.41) and (184.83,130.43) .. (184.84,129.22) -- cycle ;
\draw  [color={rgb, 255:red, 0; green, 0; blue, 0 }  ,draw opacity=1 ][fill={rgb, 255:red, 0; green, 0; blue, 0 }  ,fill opacity=1 ][line width=1.5]  (263.33,158.07) .. controls (263.34,156.86) and (264.4,155.9) .. (265.7,155.91) .. controls (266.99,155.93) and (268.02,156.91) .. (268.01,158.12) .. controls (267.99,159.32) and (266.93,160.29) .. (265.64,160.27) .. controls (264.35,160.26) and (263.31,159.27) .. (263.33,158.07) -- cycle ;
\draw  [color={rgb, 255:red, 0; green, 0; blue, 0 }  ,draw opacity=1 ][fill={rgb, 255:red, 0; green, 0; blue, 0 }  ,fill opacity=1 ][line width=1.5]  (374.94,129.22) .. controls (374.95,128.02) and (376.01,127.05) .. (377.3,127.07) .. controls (378.59,127.08) and (379.63,128.07) .. (379.62,129.27) .. controls (379.6,130.48) and (378.54,131.44) .. (377.25,131.43) .. controls (375.96,131.41) and (374.92,130.43) .. (374.94,129.22) -- cycle ;
\draw [line width=1.5]    (251.83,118.71) -- (202.05,79.26) ;
\draw [shift={(198.91,76.78)}, rotate = 398.39] [fill={rgb, 255:red, 0; green, 0; blue, 0 }  ][line width=0.08]  [draw opacity=0] (13.4,-6.43) -- (0,0) -- (13.4,6.44) -- (8.9,0) -- cycle    ;
\draw [line width=1.5]    (279.2,116.69) -- (355.41,66.89) ;
\draw [shift={(358.76,64.7)}, rotate = 506.84] [fill={rgb, 255:red, 0; green, 0; blue, 0 }  ][line width=0.08]  [draw opacity=0] (13.4,-6.43) -- (0,0) -- (13.4,6.44) -- (8.9,0) -- cycle    ;
\draw [line width=1.5]    (376.76,81.48) -- (377.09,117.39) ;
\draw [shift={(377.12,121.39)}, rotate = 269.48] [fill={rgb, 255:red, 0; green, 0; blue, 0 }  ][line width=0.08]  [draw opacity=0] (13.4,-6.43) -- (0,0) -- (13.4,6.44) -- (8.9,0) -- cycle    ;
\draw [line width=1.5]    (186.67,81.48) -- (186.99,117.39) ;
\draw [shift={(187.03,121.39)}, rotate = 269.48] [fill={rgb, 255:red, 0; green, 0; blue, 0 }  ][line width=0.08]  [draw opacity=0] (13.4,-6.43) -- (0,0) -- (13.4,6.44) -- (8.9,0) -- cycle    ;
\draw [line width=1.5]    (265.88,186.82) -- (265.56,167.32) ;
\draw [shift={(265.52,164.32)}, rotate = 449.08] [color={rgb, 255:red, 0; green, 0; blue, 0 }  ][line width=1.5]    (14.21,-4.28) .. controls (9.04,-1.82) and (4.3,-0.39) .. (0,0) .. controls (4.3,0.39) and (9.04,1.82) .. (14.21,4.28)   ;
\draw [line width=1.5]    (251.11,137.49) .. controls (231.76,129.12) and (225.01,164.64) .. (254.45,158.89) ;
\draw [shift={(258.31,157.95)}, rotate = 523.96] [fill={rgb, 255:red, 0; green, 0; blue, 0 }  ][line width=0.08]  [draw opacity=0] (13.4,-6.43) -- (0,0) -- (13.4,6.44) -- (8.9,0) -- cycle    ;
\draw [line width=1.5]    (273.44,157.95) .. controls (301.48,166.55) and (306.13,133.68) .. (284.92,137.28) ;
\draw [shift={(281.36,138.16)}, rotate = 342.28999999999996] [fill={rgb, 255:red, 0; green, 0; blue, 0 }  ][line width=0.08]  [draw opacity=0] (13.4,-6.43) -- (0,0) -- (13.4,6.44) -- (8.9,0) -- cycle    ;
\draw [line width=1.5]    (247.51,134.4) .. controls (217.1,128.6) and (214.22,206.67) .. (243.98,204.32) ;
\draw [shift={(247.87,203.6)}, rotate = 523.96] [fill={rgb, 255:red, 0; green, 0; blue, 0 }  ][line width=0.08]  [draw opacity=0] (13.4,-6.43) -- (0,0) -- (13.4,6.44) -- (8.9,0) -- cycle    ;
\draw [line width=1.5]    (186.31,27.21) -- (186.6,43) ;
\draw [shift={(186.67,47)}, rotate = 268.96] [fill={rgb, 255:red, 0; green, 0; blue, 0 }  ][line width=0.08]  [draw opacity=0] (13.4,-6.43) -- (0,0) -- (13.4,6.44) -- (8.9,0) -- cycle    ;
\draw [line width=1.5]    (161.83,28.55) -- (179.33,43.71) ;
\draw [shift={(182.35,46.33)}, rotate = 220.9] [fill={rgb, 255:red, 0; green, 0; blue, 0 }  ][line width=0.08]  [draw opacity=0] (13.4,-6.43) -- (0,0) -- (13.4,6.44) -- (8.9,0) -- cycle    ;
\draw [line width=1.5]    (212.95,29.22) -- (194.51,43.54) ;
\draw [shift={(191.35,45.99)}, rotate = 322.18] [fill={rgb, 255:red, 0; green, 0; blue, 0 }  ][line width=0.08]  [draw opacity=0] (13.4,-6.43) -- (0,0) -- (13.4,6.44) -- (8.9,0) -- cycle    ;
\draw [line width=1.5]    (375.68,27.21) -- (375.97,43) ;
\draw [shift={(376.04,47)}, rotate = 268.96] [fill={rgb, 255:red, 0; green, 0; blue, 0 }  ][line width=0.08]  [draw opacity=0] (13.4,-6.43) -- (0,0) -- (13.4,6.44) -- (8.9,0) -- cycle    ;
\draw [line width=1.5]    (351.2,28.55) -- (368.7,43.71) ;
\draw [shift={(371.72,46.33)}, rotate = 220.9] [fill={rgb, 255:red, 0; green, 0; blue, 0 }  ][line width=0.08]  [draw opacity=0] (13.4,-6.43) -- (0,0) -- (13.4,6.44) -- (8.9,0) -- cycle    ;
\draw [line width=1.5]    (402.32,29.22) -- (383.88,43.54) ;
\draw [shift={(380.72,45.99)}, rotate = 322.18] [fill={rgb, 255:red, 0; green, 0; blue, 0 }  ][line width=0.08]  [draw opacity=0] (13.4,-6.43) -- (0,0) -- (13.4,6.44) -- (8.9,0) -- cycle    ;
\draw [line width=1.5]    (100.19,129.59) -- (117.43,129.35) ;
\draw [shift={(121.43,129.29)}, rotate = 539.2] [fill={rgb, 255:red, 0; green, 0; blue, 0 }  ][line width=0.08]  [draw opacity=0] (13.4,-6.43) -- (0,0) -- (13.4,6.44) -- (8.9,0) -- cycle    ;
\draw [line width=1.5]    (101.58,152.4) -- (117.87,136.14) ;
\draw [shift={(120.7,133.31)}, rotate = 495.05] [fill={rgb, 255:red, 0; green, 0; blue, 0 }  ][line width=0.08]  [draw opacity=0] (13.4,-6.43) -- (0,0) -- (13.4,6.44) -- (8.9,0) -- cycle    ;
\draw [line width=1.5]    (102.4,104.77) -- (117.7,121.94) ;
\draw [shift={(120.36,124.93)}, rotate = 228.3] [fill={rgb, 255:red, 0; green, 0; blue, 0 }  ][line width=0.08]  [draw opacity=0] (13.4,-6.43) -- (0,0) -- (13.4,6.44) -- (8.9,0) -- cycle    ;
\draw [line width=1.5]    (266.95,240.55) -- (266.67,224.76) ;
\draw [shift={(266.6,220.76)}, rotate = 448.99] [fill={rgb, 255:red, 0; green, 0; blue, 0 }  ][line width=0.08]  [draw opacity=0] (13.4,-6.43) -- (0,0) -- (13.4,6.44) -- (8.9,0) -- cycle    ;
\draw [line width=1.5]    (291.43,239.22) -- (273.94,224.05) ;
\draw [shift={(270.92,221.43)}, rotate = 400.93] [fill={rgb, 255:red, 0; green, 0; blue, 0 }  ][line width=0.08]  [draw opacity=0] (13.4,-6.43) -- (0,0) -- (13.4,6.44) -- (8.9,0) -- cycle    ;
\draw [line width=1.5]    (240.31,238.52) -- (258.76,224.21) ;
\draw [shift={(261.92,221.76)}, rotate = 502.2] [fill={rgb, 255:red, 0; green, 0; blue, 0 }  ][line width=0.08]  [draw opacity=0] (13.4,-6.43) -- (0,0) -- (13.4,6.44) -- (8.9,0) -- cycle    ;
\draw (86.69,126.49) node    {$x_{t}$};
\draw (187.03,16.21) node    {$x_{t}$};
\draw (377.12,16.21) node    {$x_{t}$};
\draw (268.4,247.33) node    {$x_{t}$};
\draw (265.88,203.6) node    {$f_{t}$};
\draw (266.24,128.36) node    {$c_{t}$};
\draw (186.67,64.7) node    {$i_{t}$};
\draw (377.48,64.3) node    {$o_{t}$};
\draw (142.03,127.09) node  [font=\footnotesize]  {$\tanh$};
\draw (323.48,128.03) node  [font=\footnotesize]  {$\tanh$};
\draw (247.32,68.32) node  [font=\footnotesize] [align=left] {{\small {\fontfamily{pcr}\selectfont Input gate}}};
\draw (305.2,56.17) node  [font=\footnotesize] [align=left] {{\small {\fontfamily{pcr}\selectfont Output gate}}};
\draw (335.91,203.52) node  [font=\footnotesize] [align=left] {{\small {\fontfamily{pcr}\selectfont Forget gate}}};
\draw (265.52,99.4) node   [align=left] {{\footnotesize {\fontfamily{pcr}\selectfont Cell}}};
\draw (430.6,127.09) node    {$h_{t}$};
\end{tikzpicture}
\caption{Long Short-term Memory Unit}
\label{fig:lstm}
\end{figure}
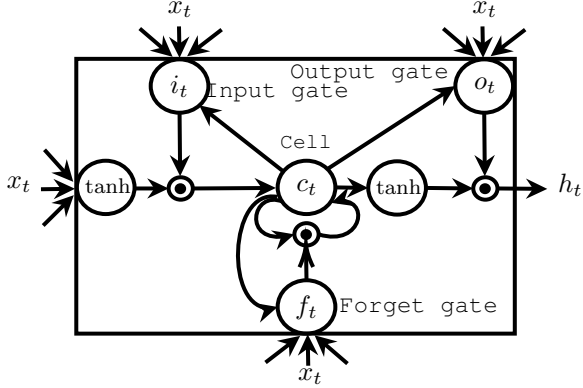

In practice, vanilla RNN encounters numerical computation difficulties. One reason presented in~\cite{bengio1994learning} is that it would cause the gradient to vanish and explode while computing the back-propagation through time, on data with long term dependencies. The vanilla RNNs only consider short term dependencies. The Long Short-Term Memory (LSTM) technique was therefore introduced to mitigate this kind of risk. The latter incorporate a memory cell~$c$ together with an input gate~$i$, an output gate~$o$ and a forget gate~$f$. The memory cell enables the network to remember its state over time, and by doing so it is
possible for the full network to capture long-term temporal
dependencies present in the training data. The evolution of LSTM
states are determined by: 
\begin{equation}
    i_{t} = \sigma(W_{i}x_{t} + U_{i}h_{t-1} + V_{i}c_{t-1} + b_{i})
\end{equation}%
\begin{equation}
    f_{t} = \sigma(W_{f}x_{t} + U_{f}h_{t-1} + V_{f}c_{t-1} + b_{f} )
\end{equation}
\begin{equation}
    c_{t} = f_{t}\odot c_{t-1} + i_{t}\odot \tanh(W_{c}x_{t} + U_{c}h_{t-1} + b_{c})
\end{equation}
\begin{equation}
    o_{t} = \sigma(W_{o}x_{t} + U_{o}h_{t-1} + V_{o}c_{t-1} + b_{o})
\end{equation}
\begin{equation}
    h_{t} = o_{t}\odot \tanh(c_{t})
\end{equation}
where $W_{\cdot},V_{\cdot}, U_{\cdot}$ and $b_{\cdot}$ are learnable parameters. The function $\sigma$ denotes sigmoid activation function, that is applied element-wise. The quantities $i_{t}$, $f_{t}$, and $o_{t}$ stand for the input, forget and output gates respectively. The output of the LSTM cell is $o_{t}$ and $\odot$
denoting point-wise vector products, i.e., Hadamard product. Fig.~\ref{fig:lstm} illustrates the learning mechanism through the LSTM cell.

\subsubsection{Recurrent Conditional Generative Adversarial Networks} 

\begin{figure} 
\centering
\includegraphics[width=0.55\linewidth,angle=270]{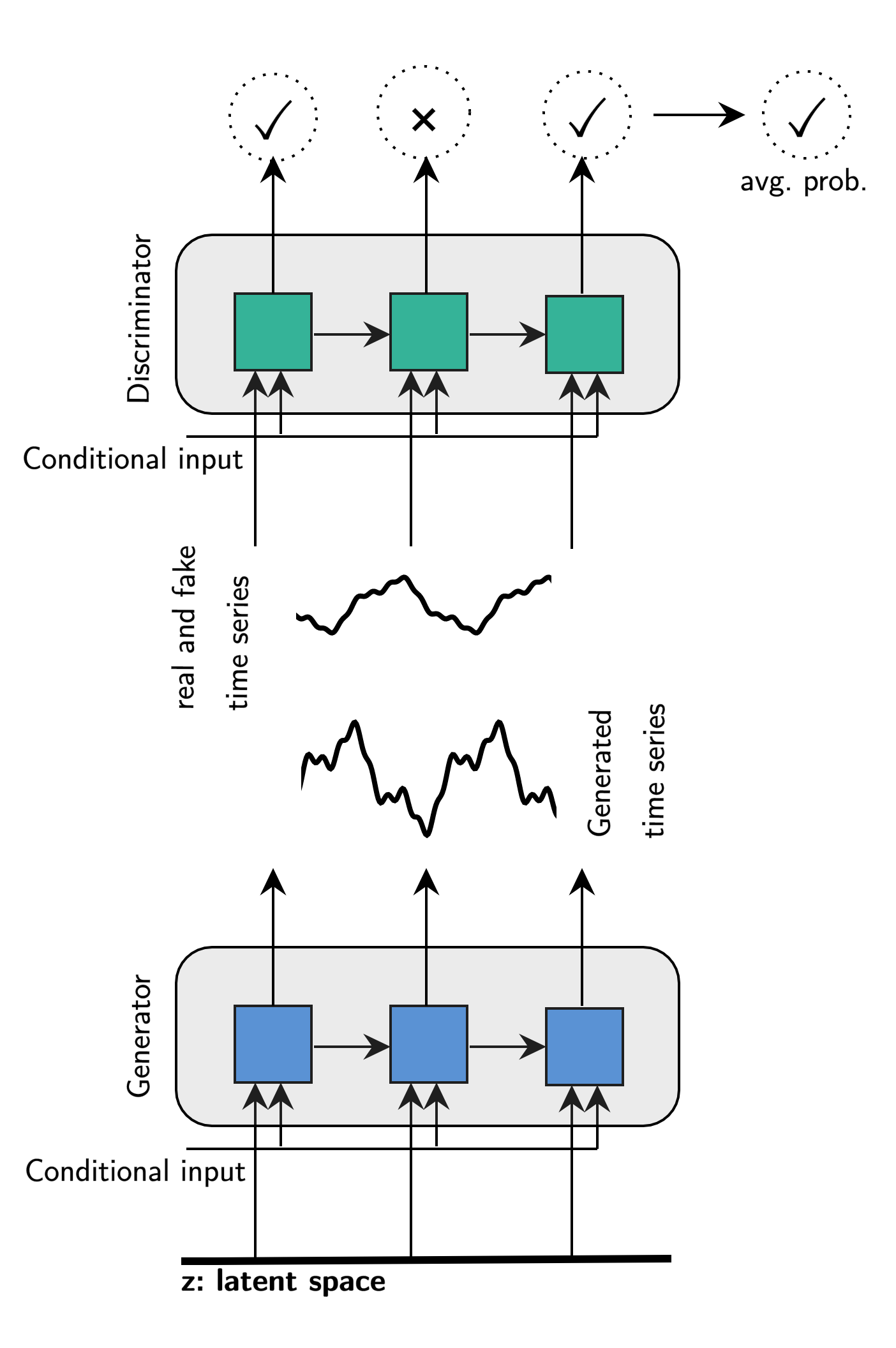}
\caption{Architecture of RCGAN composed of an RNN generator (bottom) and RNN discriminator (top). Both RNNs in the generator and discriminator, respectively, are LSTM based. The generator takes input from a latent space as well as a conditional input at each time frame. The discriminator takes either a real or fake time series together with the conditional input as input at each time frame.}
\label{fig:rcgan} 
\end{figure}

RCGANs are  generative recurrent neural networks that aim at generating real-valued time series subject to a conditional information. In the RCGAN architecture there are two different LSTM-RNNs trained simultaneously, a generator $G$ and a discriminator $D$, which have conflicting objectives. The generator learns over the training data, whereas the goal of the discriminator is to discriminate between the synthetic data generated by $G$ and the real data, as depicted in Fig.~\ref{fig:rcgan}. We denote by $k, l$ and $m$ the feature dimensions of the data, the conditional information and the latent/noise space, respectively.  Let $L$ be the length of the time series and $x_{t} \in \R^{L \times k}, y_{t} \in \R^{L \times l}, z_{t} \in \R^{L \times m}$.

In practice, the min-max game problem is described as: 
\begin{equation}
\begin{split}
\min_{G}\max_{D} \E_{x \sim p_\text{real}(x)}[\log(D(x|y)] \\ + \E_{z \sim p_{z}(z)}[\log(1-D(G(z|y))]
\end{split}
\end{equation}
where $p_\text{real}$ is the distribution of the real data and $p_{z}$ is a prior distribution over the
input noise variables. These latter, i.e.,  the $m$ sequences of $L$ points, are drawn independently from $\mathcal{N}(0,1)$.

In our case, the input consists of five-dimensional time series data, i.e., the signals for $X=60$ trips, with a binary condition attributing the type of driving (normal/aggressive). The length of all time series is equal to $60$. For more details about the architecture of our trained RCGAN, see Table I. Our RCGAN generates an example from a specific class. In other words, if we ask for the aggressive class, the generator produces one aggressive trip. Thus, after training the model, the number of generated trips per each class, should be defined. We fed the RCGAN with the training set, and then generated $0.5X, 1X,$ and $1.5X$ new trips from the data.

\subsection{Data evaluation}

In order to evaluate the quality of the RCGAN model, we used a semi-supervised framework to classify whether a trip is aggressive or normal. Firstly, we extracted statistical features from the real and fake data. Nine statistical features were calculated out of the five time series to measure different properties of that variable, namely: mean, median, mode, standard deviation, skewness, kurtosis, 25 percentile, 75 percentile and interquartile range. A further description of a few of the statistical features is given below.

We denote by $t = (t_{1},...,t_{L})$ a real-valued series with $\mu$ and $\sigma$ its mean and standard deviation, respectively. 
\subsubsection{Mode}
The mode is the most frequently appeared value in the serie.
\subsubsection{Skewness}
Skewness is used to measure the asymmetry of the data. Let $\mu^{(n)}$ be the $n$th moment, i.e.,
\begin{equation}
    \mu^{(n)}=\frac{1}{L}\sum_{i=1}^{L}{(t_{i}-\mu)^n}
\end{equation}
The skewness is then calculated with the third moment as $\text{s}=\frac{\mu^{(3)}}{\sigma^{3}}$.

\subsubsection{Kurtosis}
Kurtosis is used to measure the peakedness of the probability distribution of the data and calculated as $k=\frac{\mu^{(4)}}{\sigma^{4}}$, where $\mu_{4}$ is the $4^{th}$ moment


\subsubsection{Percentiles}
A percentile is the value of a variable below which a
certain percent of observations fall. In other words, the $p^{th}$
percentile is a value $l$ such that at most $(100\times p)\%$ of the measurements are less than this value and $100\times(1-p)\%$ are greater.
\subsubsection{Interquartile Range}
Interquartile Range (IQR) is a measure of statistical dispersion. It is defined as the difference between the $75^{th}$ and the $25^{th}$ percentiles, called the upper and lower quartiles.

The unlabeled part of the dataset were used for training an
Autoencoder (AE), see Fig. 1, in order to transfer its weights and  biases to the DNN classifier. The Autoencoder is a neural network, which aims to reconstruct the input, i.e., the target output is the input. It is composed of two main parts, an encoder that serves to compress the data in a lower dimensional space and a decoder which reproduces the input out of the bottleneck. The AE is trained in order to minimize the error between the real input and the constructed one.
More formally, let $s$ be the input, where $s \in \R^d$, the compressed representation $q \in \R^p$, $(p<d)$ mapped by $g_{\theta}$, 
\begin{equation}
    q = g_{\theta}(s) = \sigma(Ws + b) \,,
\end{equation}
where $\sigma$, $W$, and $b$, are respectively the activation function
of the encoder, the weight matrix, and the bias vector. The function $g_{\theta}$ is parameterized by $\theta = \{W,b\}$.
The decoder part
reconstructs the input from the hidden representation $q$ by the function $f_{\phi}$,
\begin{equation}
    s' = f_{\phi}(q) = \sigma(W'q + b') \,,
\end{equation}
where $\sigma$, $W'$, and $b'$ are respectively the activation function of the decoder, the weight matrix, and the bias vector. $f_{\phi}$ is parameterized by $\phi = \{W',b'\}$.

Each training input vector $s^{(i)}$ is mapped to a corresponding $q^{(i)}$ which is then mapped to a reconstruction $s'^{(i)}$ such that $s^{(i)} \approx s'^{(i)}$.
The parameters $\theta$ and $\phi$ of the model are optimized to minimize the average reconstruction error such that
\begin{equation}
\begin{split}
    (\theta^{*},\phi^{*})= \operatorname*{argmin}_{\theta,\phi}{\frac{1}{n}\sum_{i=1}^{n}{L(s^{(i)},s'^{(i)})}} \\
    = \operatorname*{argmin}_{\theta,\phi}{\frac{1}{n}\sum_{i=1}^{n}{L(s^{(i)},f_{\phi}{(g_{\theta}(s^{(i)}))})}}  \,,
\end{split}
\end{equation}
with $L$ the loss function and is given by $L(x,y) = \left\|x-y\right\|^2_{2}$.

After training the Autoencoder on the unlabeled dataset, i.e., the $178$ trips from the simulator that was not used to train RCGAN, we use the weights and biases to initialize a supervised deep neural network (DNN) model and then fine-tune the DNN model using the labeled dataset to classify the type of driving.
To measure how generated data can improve the data classification, we run various groups of experiments. In the first group which is our baseline, the classifier was
trained and validated using only the real labeled dataset.
In the following groups, we made all the combination of the training and the validation sets containing labeled real/fake/real+fake datasets.
All the classifiers were trained only with the selected
features. 

We evaluate the classifier's performance by measuring the Area Under Receiver Operating Characteristic (AUROC). This criterion is one of the most widely used metric to score the goodness of a predictor in a binary classification task. It ranges in value from $0$ to $1$. The higher the AUROC, the better the classifier is at predicting the classes, which is the type of driving in our case. The AUROC is computed on the test set containing all the real data.


\section{Results and Discussion}
\begin{figure} 
\centering
Accelerations for a normal real trip
\includegraphics[width=0.95\linewidth]{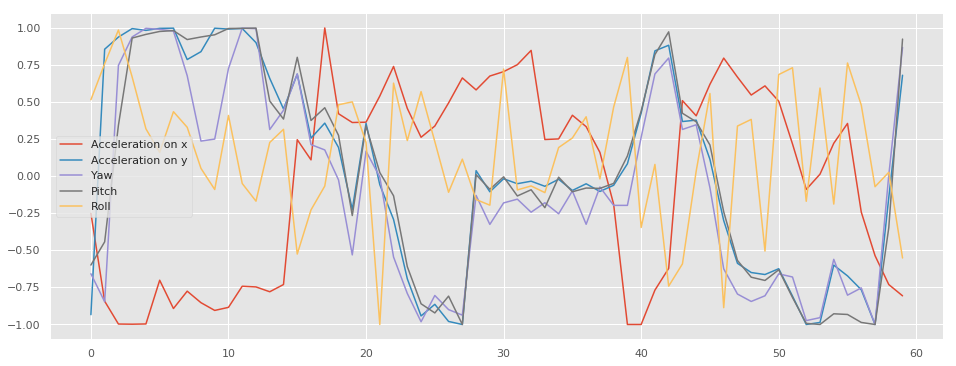}

Accelerations for a normal fake trip
\includegraphics[width=0.95\linewidth]{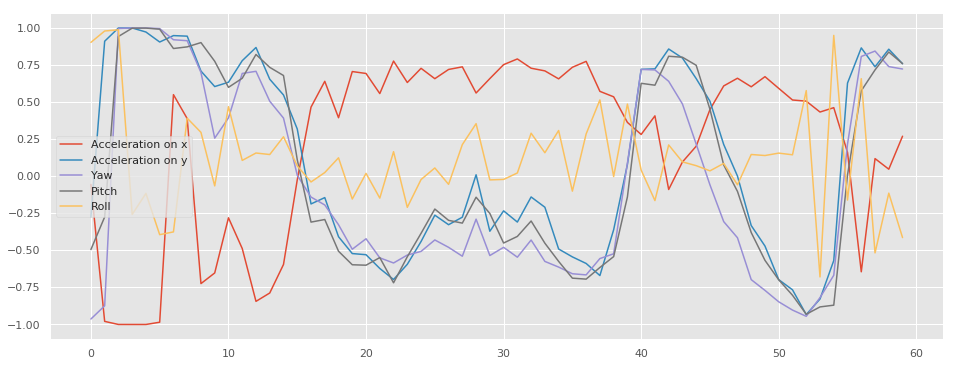}
\caption{Longitudinal acceleration (red), lateral acceleration (blue), yaw (purple), pitch (grey) and roll (yellow) are plotted for both a normal real and fake trips.}
  \label{fig:seriesexample} 
\end{figure}

Fig.~\ref{fig:seriesexample} depicts a recorded trip and a generated trip, both labeled as normal. The figure illustrates how the RCGAN was able to grasp the correlation between the different signals of a normal trip, as well as the main patterns. 
In order to investigate the quality
of the generated fake data and see whether it can be useful on a
practical level, we applied our semi-supervised framework as
an extrinsic evaluation. The generated fake data were used
in both the training and the validation set of the
classifier. All combinations of real and generated fake data is covered in Table~II.

We ran the experiments 200 times. After each trained RCGAN, we generated different amount of data and we utilized them in the validation or training set of our classifier.
In $79\%$ of the simulations the RCGAN reached at least
an AUROC strictly higher than the baseline value, for at least one combination of real and fake data in every of the 200 runs. Table~II shows the performance of the classifier of the semi-supervised framework, trained and validated on different sets consisting of combinations of real and fake data for a simulation outperforming the baseline. AUROC is measured on the test set which contains all the real trips. Bold depicts the AUROC superior to the baseline values.
We can see that for most simulations the AUROC exceeds the baseline, for a variety of sets and ratios of real and generated fake data. 

\begin{table}
\centering
\caption{Parameters for the three neural networks}
\begin{threeparttable}
\begin{tabular}{ll} \hline
\textbf{RCGAN} \\ 
Learning rate & 0.001 \\
Batch size & 1 \\
Number of epochs & 5000 \\
Generator optimizer & ADAM \\
Discriminator optimizer & Gradient Descent \\
Generator rounds & 1 \\
Discriminator rounds & 1 \\
RNNs hidden units & 100 \\
Latent dimensions & 25 \\
Smooth rate & 0.1 \\[0.5em]
\textbf{Autoencoder} & \\
Number of epochs & 100 \\
Hidden layers & [100,50,100] \\
Activation function & $\tanh$ \\[0.5em]

\textbf{Classifier} & \\
 Hidden layers & [100,50,100] \\
Learning rate\tnote{1} & 0.001, 0.01, 0.1 \\
 Number of epochs\tnote{1} & 100, 200, 500 \\
Activation function\tnote{1}& tanh, maxout, rectifier \\ 
\end{tabular}

\label{tab:balance222}
\begin{tablenotes}
\item[1]Grid search was based on these parameters for the 
\end{tablenotes}
\end{threeparttable}
\end{table}

Since we varied the percentage of real and generated fake data in both training and validation sets of the classifier, it is of interest of how much generated fake data that is needed and how it should be utilized by the classifier. Table~III highlights the summary over the set of simulations which outperform the baseline, i.e., the number of recorded AUROCs that exceeds the baseline value, for each combination set and ratio fake.

\begin{table}
\centering
\caption{Performance of the classifier}
\footnotesize
\begin{threeparttable}
\begin{tabular}{ cccc } \hline
Training Set & Validation Set & Ratio Fake/Real  & AUROC\tnote{1}\\ \hline
\multirow{1}{*}{R\tnote{2}} & \multirow{1}{*}{R}  & 0\% & 0.823\\ \hline\hline
\multirow{3}{*}{R + F\tnote{3}} & \multirow{3}{*}{R + F}   & 50\% & \textbf{0.858} \\
 & & 100\% & \textbf{0.851} \\
 & & 150\% & 0.805 \\\hline
\multirow{3}{*}{R} & \multirow{3}{*}{F}  &  50\% & \textbf{0.846} \\
& & 100\% & 0.774 \\
& & 150\% & \textbf{0.845} \\\hline
\multirow{3}{*}{F} & \multirow{3}{*}{R} &  50\% & 0.799 \\
& & 100\% & \textbf{0.858}\\
& & 150\% & \textbf{0.851}\\ \hline
\multirow{3}{*}{R + F} & \multirow{3}{*}{R}  & 50\% & \textbf{0.851} \\
& & 100\% &  \textbf{0.832}\\
& & 150\% &  \textbf{0.844}\\ \hline
\multirow{3}{*}{F}  &\multirow{3}{*}{F}  &  50\% & 0.776 \\
 & & 100\% & \textbf{0.846} \\
 & & 150\% & \textbf{0.833} \\ \hline
\multirow{3}{*}{R} & \multirow{3}{*}{R + F} &  50\% & \textbf{0.841} \\
 & & 100\% & \textbf{0.851} \\
 & & 150\% & 0.813 \\ \hline
\multirow{3}{*}{F} & \multirow{3}{*}{R + F}  &  50\% & \textbf{0.841} \\
 & & 100\% & 0.805 \\
 & & 150\% & 0.805 \\ \hline
\multirow{3}{*}{R + F} & \multirow{3}{*}{F} & 50\% & \textbf{0.849} \\
 & & 100\% & 0.805 \\
 & & 150\% & \textbf{0.841} \\
\hline
\end{tabular}
\label{tab:balance}
\begin{tablenotes}
\item[1]This measure is computed on the test set, containing all the real data, namely the $238$ trips.
\item[2]The set R consists of the $60$ real trips which were used to train the RCGAN.
\item[3]The set F consists of the fake data which were generated from the RCGAN.

\end{tablenotes}
\end{threeparttable} 
\end{table}
On a first glance, we can divide the Table~III into three groups. First one containing the combination set which have the highest total number. Training on the real, while validating on both the real and the fake seems to be the best option in order to ensure a better classification.
\begin{table}
\centering
\caption{set of simulations which outperform the baseline}
\begin{threeparttable}
\footnotesize
\begin{tabular}[b]{c c c c c c c}
\hline
\multicolumn{2}{c}{Sets} & \multicolumn{3}{c}{Counts\tnote{1}} &
\multicolumn{2}{c}{AUROC} \\
\hline\hline
Training & Validation & 50\% & 100\% & 150\% & Mean & SD\\
\hline
R+F & R+F & 21  & 14  & 14 & 0.834 & 0.009\\
R & F &  76  & 76  & 86  & 0.843 &0.007\\
F & R & 3  &  1   & 2 & 0.833& 0.004 \\
R+F & R & 24 &  14 &  15 & 0.833 &0.008 \\
F & F & 0  &  1  &  0 & 0.830 & -- \\
R & R+F & 106 & 101 & 104 & 0.840 & 0.008\\
F & R+F & 2 &   0 &   1 & 0.835 & 0.003 \\
R+F & F & 12 &  26  & 17 &0.835& 0.009 \\
\hline
\end{tabular}
\label{tab:sam_count2}
\begin{tablenotes}
\item[1] The recorded number for each sets and percentage of fake data.
\end{tablenotes}
\end{threeparttable}
\end{table}

Training on the real data and validating only on the fake data can be also a good way to use the generated data. The second group contains the following combinations; training on the real data and fake data, while validating on the real data, training and validating on both the fake data and real data, and lastly training on real data and fake data while validating on the fake data. This group is characterised by a lower number of records comparing to the first one. This underlines the fact that incorporating the fake data in the training set is less likely to improve the classifier. The third group contains the remaining combinations. This group is characterised by the fact that the training set is only composed of fake data. The negligible number of this group excludes the possibility of using only the fake data to improve the classifier accuracy. 
This result can be justified by the fact that the generation of data is done on the basis of the real ones, therefore substituting the content of the classifier's training set from real data to fake data, would not guarantee an improvement. The generative model had to learn from the real data to end up having new ones close enough to the original, but still different.

\begin{figure} 
    \centering
   \includegraphics[width=0.9\linewidth]{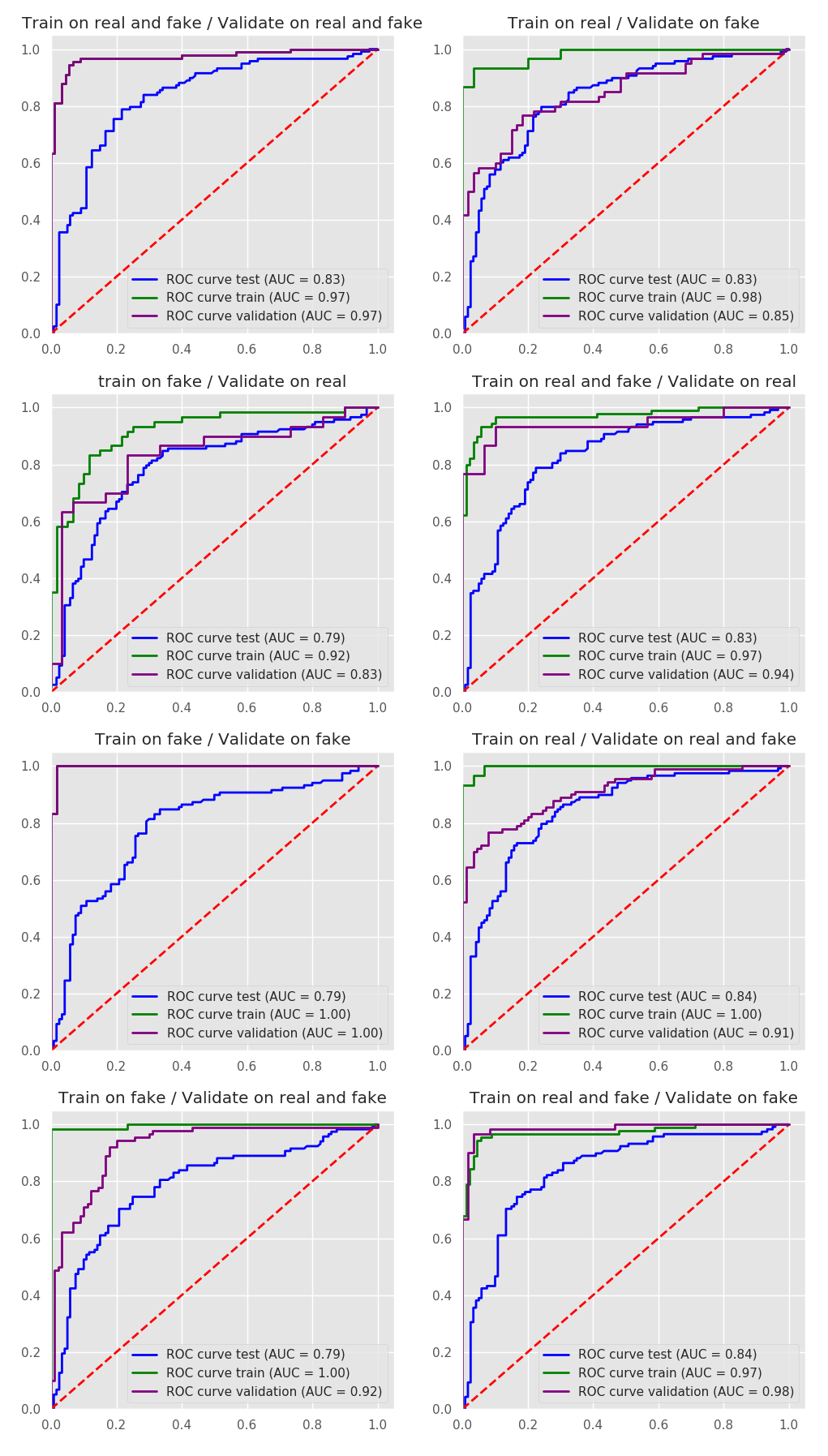}
  \caption{The eight plots denote the ROC curve on the training, validation and test set for the classifier with a ratio of Fake on Real equals to $100\%$.
  The test set contains all the real data. The $x$-axis and $y$-axis depict the False Positive Rate and the True Positive Rate, respectively.}
  \label{fig:roc} 
\end{figure}

Fig.~\ref{fig:roc} shows that the classifier can perform well by training merely on the generated fake data. By training and validating on the fake data, we can have an AUROC slightly lower than the baseline, which still guarantees a good prediction of the type of driving.

The first group also reached the highest average of AUROC between its elements, comparing to the other ones.
Consequently, we capture the importance of incorporating the fake data only into the validation set.

On the other hand, we want to see whether the size of the generated data would affect the performance of the classifier. Since we know from the previous results, in which combination sets the fake data worth to be used, we limited the scope only on the first group, which only train the classifier on the real data. We can see in Table III, that increasing a ratio fake to $150\%$ would give in overall, higher chances to improve the model. In this case, it means that synthesising more data than the size of the original one, can give a better classification of driving behavior.

\section{Conclusion}
In this paper, we outlined our experiences of using Recurrent Conditional GANs for generating IMU signals, which are assessed by the improvement of a semi-supervised framework to classify the type of driving. The classification applied on the extracted features of the real and synthetic data, was mostly improved by using the latter in the validation set. 
The two main contributions in this work are the generation of IMU signals and the quantitative extrinsic assessment of the synthetic data using a deep learning based approach. 
For future research, we plan tox investigate how the parameters of the RCGAN can be improved with the aim to find the most convenient network architecture to ensure an close to optimal classification given the limited amount of labeled data.


\bibliographystyle{IEEEtran}
\bibliography{Biblio}

\end{document}